\documentclass[%
 reprint,
superscriptaddress,
 amsmath,amssymb,
 aps,
 prl
]{revtex4-2}

\usepackage{graphicx}
\usepackage{dcolumn}
\usepackage{bm,upgreek}
\usepackage{soul}
\usepackage[normalem]{ulem}
\usepackage{color}
\usepackage{physics}
\usepackage{hyperref}
\usepackage{bibunits}

\begin{document}

\newcommand{\papertitle}{From classical to quantum loss of light coherence}
\title{ \papertitle}

\author{Pierre Lass\`egues}
\affiliation{Universit\'e C\^ote d'Azur, CNRS, INPHYNI, France}

\author{Mateus Antônio Fernandes Biscassi}
\affiliation{Universit\'e C\^ote d'Azur, CNRS, INPHYNI, France}
\affiliation{Departamento de F\'{\i}sica, Universidade Federal de S\~{a}o Carlos, Rodovia Washington Lu\'{\i}s, km 235 - SP-310, 13565-905 S\~{a}o Carlos, SP, Brazil}

\author{Martial Morisse}
\affiliation{Universit\'e C\^ote d'Azur, CNRS, INPHYNI, France}

\author{Andr\'e Cidrim}
\affiliation{Departamento de F\'{\i}sica, Universidade Federal de S\~{a}o Carlos, Rodovia Washington Lu\'{\i}s, km 235 - SP-310, 13565-905 S\~{a}o Carlos, SP, Brazil}
\affiliation{Department of Physics, Stockholm University, 10691 Stockholm, Sweden}

\author{Pablo Gabriel Santos Dias}
\affiliation{Departamento de F\'{\i}sica, Universidade Federal de S\~{a}o Carlos, Rodovia Washington Lu\'{\i}s, km 235 - SP-310, 13565-905 S\~{a}o Carlos, SP, Brazil}

\author{Hodei Eneriz}
\affiliation{Universit\'e C\^ote d'Azur, CNRS, INPHYNI, France}

\author{Raul Celistrino Teixeira}
\affiliation{Departamento de F\'{\i}sica, Universidade Federal de S\~{a}o Carlos, Rodovia Washington Lu\'{\i}s, km 235 - SP-310, 13565-905 S\~{a}o Carlos, SP, Brazil}

\author{Robin Kaiser}
\affiliation{Universit\'e C\^ote d'Azur, CNRS, INPHYNI, France}

\author{Romain Bachelard}
\email{romain@ufscar.br}
\affiliation{Universit\'e C\^ote d'Azur, CNRS, INPHYNI, France}
\affiliation{Departamento de F\'{\i}sica, Universidade Federal de S\~{a}o Carlos, Rodovia Washington Lu\'{\i}s, km 235 - SP-310, 13565-905 S\~{a}o Carlos, SP, Brazil}

\author{Mathilde Hugbart}
\email{mathilde.hugbart@inphyni.cnrs.fr}
\affiliation{Universit\'e C\^ote d'Azur, CNRS, INPHYNI, France}

\date{\today}

\begin{abstract}
Light is a precious tool to probe matter, as it captures microscopic and macroscopic information on the system. We here report on the transition from a thermal (classical) to a spontaneous emission (quantum) mechanism for the loss of light coherence from a macroscopic atomic cloud. The coherence is probed by intensity-intensity correlation measurements realized on the light scattered by the atomic sample, and the transition is explored by tuning the balance between thermal coherence loss and spontaneous emission via the pump strength. Our results illustrate the potential of cold atom setups to investigate the classical-to-quantum transition in macroscopic systems.
\end{abstract}

\maketitle


{\em Introduction.---}Quantum mechanics has brought a completely new description of a physical system, introducing the possibility of ``entanglement'' between its different states. However, this diversity in possible states comes at the expense of a dramatic increase in complexity as one aims at an exhaustive description of the system. To compensate for the exponential growth of the associated Hilbert space with the number of constituents, one may derive an effective dynamics for a selected set of degrees of freedom, tracing over the less relevant ones. This loss of information leads to the notion of decoherence~\cite{Breuer2007}, and the partial knowledge of the system state allows for an accurate prediction of the dynamics over a finite time only. From a fundamental point of view, decoherence actually questions the notions of measurement, collapse of the wavefunction~\cite{Zurek03,Wallace2012}, and hidden variables in quantum mechanics~\cite{Zeh1970}. Looking toward quantum technologies, decoherence is a major obstacle to the preservation of quantum information, but it is also a central mechanism behind the quantum random number generation~\cite{Herrero2017}.

Let us consider the prototypical example of spontaneous emission (with rate $\Gamma$) for a quantum emitter: It arises from tracing over the electromagnetic modes in which the particle excitation may be emitted. Yet, while half of the spontaneous emission rate can be explained by the radiation reaction with a classical approach, the other half was shown to stem from the quantum fluctuations of the modes: ``\textit{Die spontane Emission ist somit eine durch die Nullpunktsschwingungen des leeren Raumes erzwungene Emission eines Lichtquants.''}, as wrote Weisskopf~\cite{Weisskopf1935}\footnote{It can be translated as ``\textit{Spontaneous emission is thus a stimulated emission of one quantum of light caused by the zero point fluctuations of vacuum.}''.}. These zero-point fluctuations do not result from a set of unknown (or ``hidden'') variables, as in classical statistical physics when microscopic details are ignored, but rather from Heisenberg's uncertainty principle~\cite{Newing1935}.

In the case of a quantum emitter, the decoherence mechanism incarnated by spontaneous emission leaves its mark on the radiated light, since signatures of the quantum nature of the emitter, such as photon antibunching~\cite{Kimble1977,Dagenais1978} or  Rabi oscillations~\cite{Shore1993,Knight1980}, are visible on a time scale $1/\Gamma$. When moving to many emitters, the nature of the mechanism at the origin of the light coherence loss can be more ambiguous, as one meets the frontier between quantum physics and statistical ensembles. For example, photon antibunching is observed in large systems under specific conditions such as phase matching~\cite{Grangier1986} or confinement of light in fibers~\cite{Prasad2020}; spontaneous emission then sets the time scale of the light coherence. Differently, the reduction of this coherence time due to the particles' motion can be understood from a classical perspective: a macroscopic information (the velocity distribution) is extracted from the reduction of the light coherence, without the knowledge of the microscopic trajectories, and this effect is at the core of the diffusive wave spectroscopy technique~\cite{Maret_1987,Pine_1988,Weitz1989,Fraden1990,Hebraud1997,Durduran2004,Eloy_2018}. This illustrates the variety of phenomena which compete to set a limit to the light coherence.

\begin{figure}[t!]
	\centering
	\includegraphics[width=1\columnwidth]{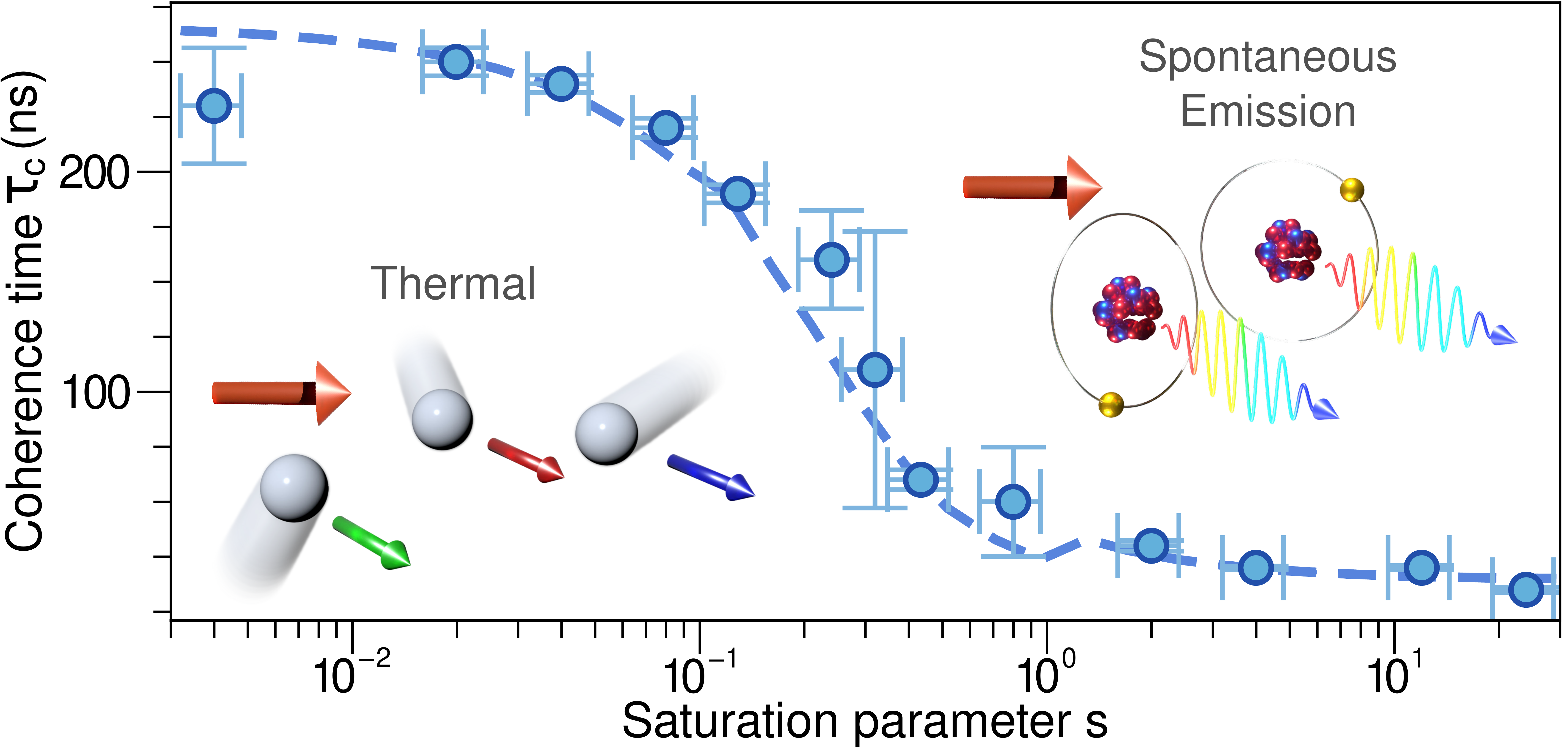}
	\caption{Coherence time of the light radiated by a macroscopic cloud, as a function of the saturation parameter (dashed: theory; dots: experiment with error bars, see main text). Left inset: For a weak monochromatic drive on the atoms, the elastically scattered light acquires a frequency shift due to the Doppler effect. The finite temperature of the cloud broadens the light spectrum, setting the coherence time. Right inset: Strongly driven atoms each emits a broadened spectrum (Mollow triplet).}
	\label{fig:phys}
\end{figure}

In this work, we report on the transition from a classical to a quantum mechanism for the loss of coherence in the light scattered by a macroscopic atomic cloud of neutral atoms. In the weak drive regime, the atoms scatter light elastically, yet the finite cloud temperature, through the Doppler effect, induces a broadening of the spectrum: the coherence loss is here a macroscopic manifestation of the microscopic dynamics (see Fig.~\ref{fig:phys}). Differently, spontaneous emission dominates the scattering from strongly driven atoms, and the light coherence is then limited by the transition rate, $\Gamma$. In this regime, the emission of each atom is spectrally broadened~\cite{Mollow_1969}, and the reduction of coherence results from zero-point fluctuations rather than from unknown microscopic details~\cite{Milonni1984}. Experimentally, we perform intensity-intensity correlation measurements to characterize the (loss of) light coherence: the associated intensity fluctuations are shown to also arise, depending on the regime explored, from either the Doppler effect or spontaneous emission. Alternatively, we monitor field-field correlations, which confirm that the electric field coherence suffers from the same mechanism as the intensity (Siegert relation), both in the classical and in the quantum regime of coherence loss. 
 
\begin{figure}[t!]
	\centering
 	\includegraphics[width=1\columnwidth]{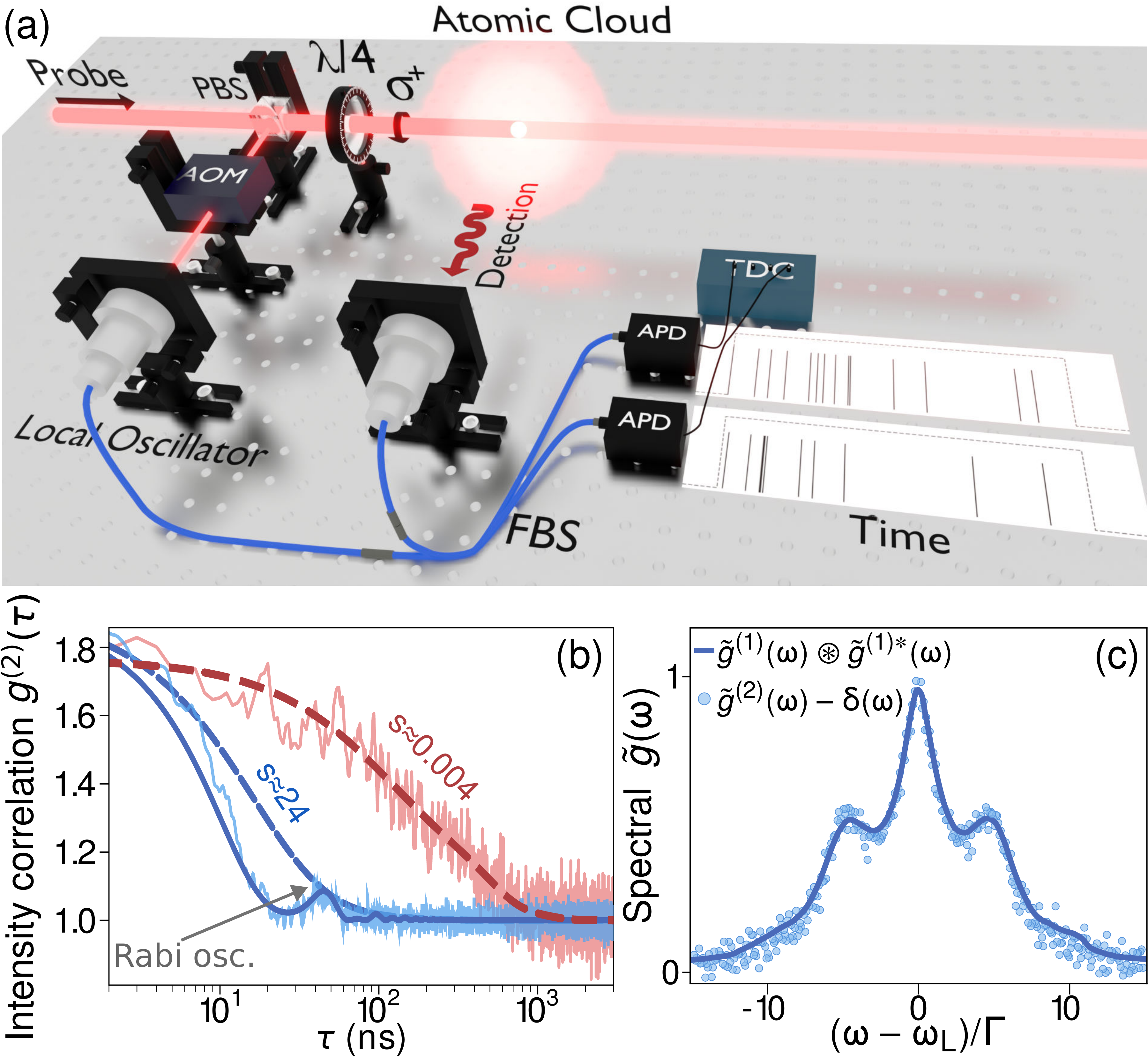}
	\caption{(a) Schematic setup of the experiment, see~\cite{SM} for details. (b) Temporal evolution of the second-order coherence $g^{(2)}(\tau)$, in a low-saturation temperature-dominated regime [$s = (4.0 \pm 0.8)\times 10^{-3}$] and in a high-saturation spontaneous emission-dominated regime ($s = 24 \pm 5$). 
	Dashed lines: fits of the decay capturing the coherence time ($\tau_c$ is computed as the half-width at half-maximum);  the solid thicker line is a fit containing the (coherent) Rabi oscillation of the saturated regime. (c) Observation of the Siegert relation, which writes $\tilde{g}^{(2)}(\omega)=\delta(\omega) + \tilde{g}^{(1)}(\omega)\circledast\tilde{g}^{(1)*}(\omega)$ in the frequency space ($\circledast$ the convolution), for the large-$s$ regime, $s\approx 60$. The elastic component is broadened by the temperature; the spectra are normalized to one.
	}
	\label{fig:setup}
\end{figure}

{\em Thermal coherence loss versus spontaneous emission.---}Our experimental setup, see Fig.~\ref{fig:setup}(a) and ~\cite{SM}, allows us to measure simultaneously the first-order (field-field) and second-order (intensity-intensity) correlation functions of the scattered light~\cite{Ferreira2020}:
\begin{eqnarray}
g^{(1)}(\tau)&=&\frac{\langle \hat{E}^-(t) \hat{E}^+(t+\tau)\rangle}{\langle \hat{E}^-(t) \hat{E}^+(t)\rangle}, \label{eq:g1}
\\ g^{(2)}(\tau)&=&\frac{\langle \hat{E}^-(t)\hat{E}^-(t+\tau) \hat{E}^+(t+\tau)\hat{E}^+(t)\rangle}{\langle \hat{E}^-(t) \hat{E}^+(t)\rangle^2}.\label{eq:g2}
\end{eqnarray}
$\hat{E}^+$ refers to the positive frequency component of the electric field in the measured mode, $\langle.\rangle$ either to the average over time or to the expectation value~\cite{Loudon:book}, and we here consider the steady-state limit, $t\to\infty$. In all cases, we also average over configurations.

Two examples of $g^{(2)}(\tau)$ correlation functions taken from the experiment are presented in Fig.~\ref{fig:setup}(b). For a weak pump, with saturation parameter $s\ll 1$, most light is scattered elastically by each atom in its own (moving) frame. In the laboratory frame, this motion translates into a change in frequency of the light (Doppler effect). The interference between the fields scattered by the disordered ensemble of moving atoms leads to a Doppler-broadened spectrum, with a coherence time $\tau_c^T\propto 1/\sqrt{T}$ ($T$ the temperature). Our temperature-induced intensity coherence time  $\tau_c^T\approx 260$\,ns, extrapolated for $s = 0$ from measurements at several low-$s$ values, corresponds to a temperature of about 200~$\upmu$K and a cloud with an optical depth of $6$~\cite{Eloy_2018}. This statistical analysis is the basis, for example, of diffusive wave spectroscopy technique~\cite{Maret_1987,Pine_1988,Weitz1989,Fraden1990,Hebraud1997,Durduran2004}, and it is a purely classical mechanism of coherence loss.

To enter the regime where coherence loss is based on quantum randomness, we use a strong resonant pump ($s\gg1$). Each atom then presents a spectrally broadened fluorescence, the so-called Mollow triplet~\cite{Mollow_1969,Ng2022}, which is characterized by a peak at resonance and two sidebands shifted by $\pm\Omega$ from the carrier. The beating between these peaks of inelastic scattering is manifested as Rabi oscillations in the $g^{(2)}(\tau)$, whereas the peak widths ($\Delta\omega\approx\Gamma$) set the coherence time of this spontaneously emitted light to $\tau_c^\textrm{SE}\approx 16$\,ns, see Fig.~\ref{fig:setup}(b). 
This broadening mechanism does not rely on the microscopic state of the system (and macroscopically captured by temperature, for example) or any ``hidden variable'', but rather on zero-point fluctuations~\cite{Milonni1984}.

In our experiment, the transition between the classical and quantum regimes occurs when the ratio of spontaneously emitted to elastically scattered power $P_\textrm{SE}/P_\textrm{ESL}$ is inverted. This ratio here corresponds to the saturation parameter $s$~\cite{Mollow_1969,Steck07}, tuned via the pump Rabi frequency $\Omega$. In Fig.~\ref{fig:phys}, we present the evolution of the coherence time of the light when crossing from the classical to the quantum regime of coherence loss. The coherence time was extracted from the second-order correlation $g^{(2)}(\tau)$ [see Fig.~\ref{fig:setup}(b)]: the curves are fitted assuming an exponential decay for the elastic component~\cite{Eloy_2018} and the Mollow triplet for the inelastic part, from which $\tau_c$ is defined as the half-width at half-maximum. The theoretical curve has been obtained by considering the radiation from a large sample of independent two-level atoms~\cite{Cohen-Tannoudji1997}. The slightly larger coherence time observed for intermediate $s$ in the experiment (see Fig.~\ref{fig:phys}), as compared to the theoretical prediction, can be attributed to the attenuation of the beam during its propagation in the cloud, which results in an increase of the relative contribution of elastic scattering. The overall agreement between the experimental and theoretical curves shows that the proposed picture of $N\gg1$ independent atoms captures well the underlying physical mechanisms of coherence loss for the light.

{\em Origin of the fluctuations.---}The finite coherence time reflects the fluctuations of the intensity over time, which can have different origins. Elastically scattered light is usually
treated as a continuously radiated field, associated with a fluctuating classical intensity because of the emitters'
motion. Although the two pictures of continuous versus discrete detection events can be reconciled~\cite{Glauber:1963b,Loudon:book}, let us now discuss how the fluctuations observed in each regime still depend on the underlying physical mechanism.

Let us first consider motionless particles scattering light elastically. One may expect the light emitted by the cloud to inherit the same statistics of the pump, $g^{(2)}(0)=1$ for a laser. Yet this classical picture presents a loophole, as one already perceives from the test case of a pair of two-level atoms. Under a resonant drive with wavevector $\mathbf{k}_L$, two remote (non-interacting) two-level atoms at positions $\mathbf{r}_{1,2}$ exhibit a second-order correlation function at zero delay which satisfies~\cite{SM,Skornia2001}:
\begin{equation}
    g^{(2)}(0)=\frac{(s+1)^2}{\left(s+1+\cos\left[(k\hat{n}-\mathbf{k}_L).(\mathbf{r}_2-\mathbf{r}_1)\right]\right)^2},\label{eq:g2N2}
\end{equation}
in the far field and in the steady state, and with $\hat{n}$ the direction of observation. The cosine is an interference term which produces an angular dependence for the $g^{(2)}(0)$. Indeed, for $N=2$ this interference term is present in the intensity $\langle \hat E^- \hat E^+\rangle$, yet absent from the two-photon term $G^{(2)}=\langle \hat E^- \hat E^- \hat E^+ \hat E^+\rangle$. This peculiarity leads to a spatial modulation of the light statistics, as observed in pairs of trapped ions~\cite{Wolf2020}. This result is at odds with a linear optics approach mentioned above, where motionless scatterers emit a constant electric field, corresponding to $g^{(2)}(0)=1$ (coherent light). The flaw in the latter approach is that the emission of two photons, as measured by the $g^{(2)}$ function, cannot be described classically. Note that this feature is absent from field-field correlations \eqref{eq:g1}, which is equal to one at zero delay by definition: the $g^{(1)}$ function does not address photons, but only fields. 

In Eq.~(\ref{eq:g2N2}), of particular interest are the angles which satisfy $(k\hat{n}-\mathbf{k}_L).(\mathbf{r}_2-\mathbf{r}_1)=\pi \mod 2\pi$. These angles correspond to destructive interference, at which the emission computed from the optical coherences of the two-level atoms (that is, the elastically scattered component) cancels. In those particular directions, only a contribution from the doubly-excited state remains and one obtains $g^{(2)}(0)=(s+1)^2/s^2$, which diverges in the $s\to 0$ limit~\cite{Skornia2001}, even in the large $N$ case~\cite{SM}.

Far from any divergence, the intensity correlations observed in the experiment leads to
$g^{(2)}(0)\approx 2$ in the low-$s$ regime, expected from chaotic light. This finite value is due to the finite temperature of the cloud, along with the finite time necessary to evaluate the $g^{(2)}(\tau)$.
Indeed, the atomic motion results in spatio-temporal fluctuations of the speckle field. Thus, the measurement of the $g^{(2)}$ over a time scale much larger than the coherence time of the speckle provides a finite {\it averaged} value of the intensity, and thus a finite intensity variance $g^{(2)}(0)$.
In other words, in the classical regime the notation $\langle\cdot\rangle$ refers to a statistical average, on the thermal probability distribution of atomic positions and velocities. 
In our experiment, the average is realized over the duration of the experiment ($>20\upmu s$), much larger than the coherence time $\tau_c^T$ of the speckle grain, and over different clouds.

\begin{figure}[t!]
	\centering
	\includegraphics[width=\columnwidth]{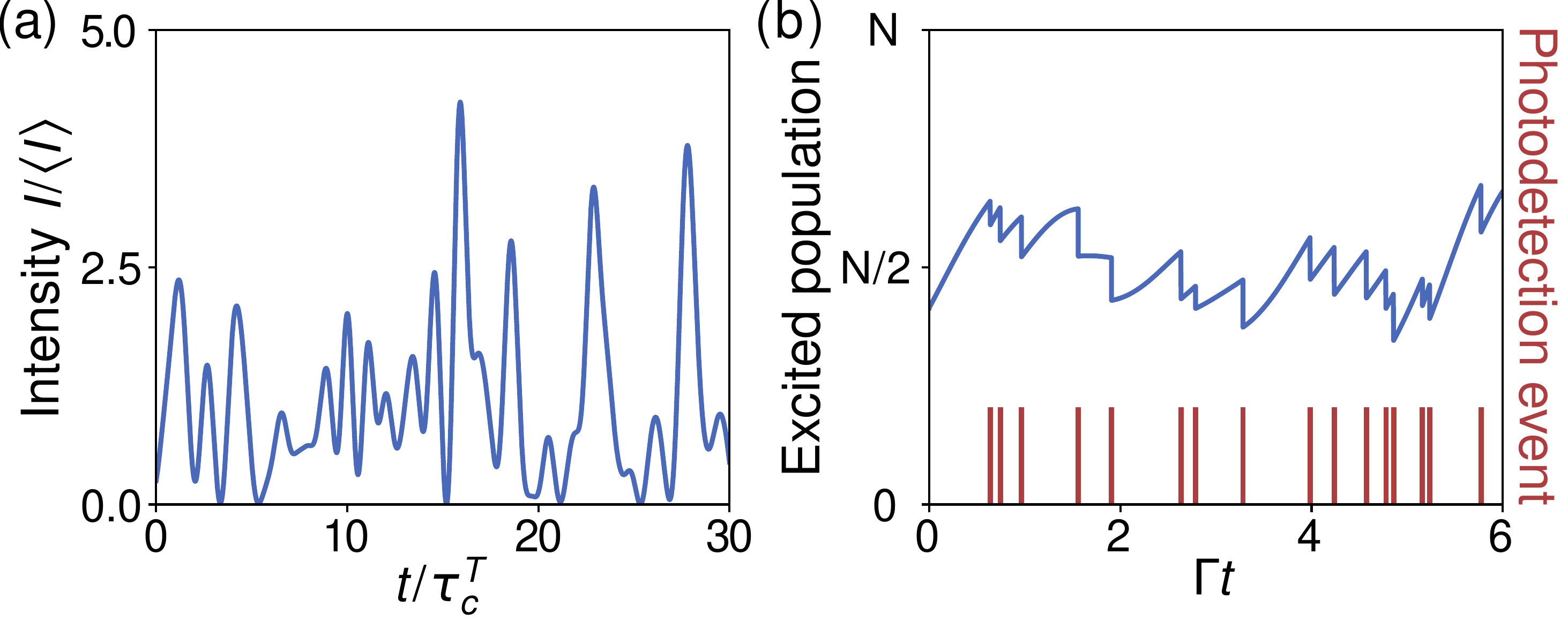}
	\caption{(a) Evolution of the emitted intensity from a cloud of moving classical dipoles with positions $\mathbf{r}_j(t)$, where the Doppler effect combines with interference to provide temporal fluctuations: $I\propto |\sum_j \exp\left[(\mathbf{k}_L-k\hat{n})\cdot\mathbf{r}_j(t)\right]|^2$, with $\hat{n}$ the direction of observation, and $\langle\dot{r}_j^2\rangle=k_BT/M$. (b) Stochastic evolution of the excited population of an ensemble of $N$ atoms driven by a strong pump, in the steady-state, obtained from exact simulations of strongly driven two-level atoms~\cite{Johansson2012,Johansson2013} (with $N=10$ and $s=8$). The quantum jumps toward a lower population state correspond to the emission of a photon (photodetection events in red).}
	\label{fig:Ievo}
\end{figure}

Although we have adopted until now a quantum-mechanical approach, resorting to two-level atoms, a description of the scatterers as moving classical dipoles leads to the same conclusion.
This is illustrated in Fig.~\ref{fig:Ievo}(a), where the intensity emitted in a given direction by classical dipoles with ballistic trajectories is shown. In particular, the value $g^{(2)}(0)= 2$ and the coherence time $\tau_c^T\approx 260$~ns obtained from this approach are the same as the ones observed in the experiment, which supports the classical origin of coherence loss in this regime.

Differently, spontaneous emission gives rise to intensity fluctuations even in absence of any motion. The random nature of the photon emission in this regime is best understood adopting the quantum jump approach~\cite{Molmer1993}.
While the atomic ensemble undergoes coherent Rabi oscillations under the action of the  drive, it also stochastically decays toward a lower population state when a photon is spontaneously emitted. This process is illustrated in Fig.~\ref{fig:Ievo}(b), where the total excited population of a cloud of $N$ atoms is presented for a given realization (or ``trajectory''). At each decay event, a photodetection event occurs (see red lines). In the experiment [inset in Fig.~\ref{fig:setup}(a)], the correlation between the detection event times of the two detectors is computed, before a binning over a time window much smaller than the coherence time of the light is applied. This experimental protocol effectively measures the expectation value, with a temporal average, of the operators involved in the definition of $g^{(2)}(\tau)$, see Eq.~\eqref{eq:g2}, and it provides the continuous curves presented in Fig.~\ref{fig:setup}(b). 

The amplitude of the fluctuations which we observe, $g^{(2)}(0)\approx 2$, is consistent with the value for chaotic light. A common picture for a gas of particles is when each emitter emits a field with a given phase, yet with a mechanism which randomizes this phase~\cite{Loudon:book}. Nonetheless, motion is unnecessary to explain the fluctuations of strongly driven atoms, since one needs to assume no other coherence loss mechanism than spontaneous emission. Indeed, saturated atoms ($s\to\infty$ limit) reach a steady-state with the fluctuation value $g^{(2)}(0)=2(1-1/N)$~\cite{SM}, where the $1/N$ term vanishes for large samples such as in our experiment. Hence, temperature, collisions or mechanisms other than spontaneous emission are not needed to interpret the fluctuations and coherence loss of the light from a large saturated atomic cloud. An additional signature that the emission in the large-$s$ regime comes from quantum emitters is the Rabi oscillation observed in the $g^{(2)}(\tau)$ of Fig.~\ref{fig:setup}(b). The oscillation results from the coherent dynamics between the two levels of the atoms (with a circularly-polarized large-$s$ pump, the atoms are driven to an extreme Zeeman sublevel, from which a two-level transition only is explored).

{\em Field fluctuations and the Siegert relation.---} Elastically scattered light has a well defined phase, which is determined by the incident laser and the trajectories of the atoms. For spontaneous emission, the broadened fluorescence spectrum of even single atoms, along with the absence of a phase operator~\cite{Noh1992,Torgerson1996}, prevent a direct analogy. Nevertheless, the electric field from spontaneous emission possesses a temporal coherence, which is captured by field-field correlations, see Eq.~\eqref{eq:g1}. In our setup, it is obtained from the homodyne measurements described previously.

In Fig.~\ref{fig:setup}(c), we present simultaneous measurements of the $g^{(1)}$ and the $g^{(2)}$ functions, in the strong drive regime ($s\approx 60$). The excellent match between intensity-intensity correlations and the square modulus of field-field correlations corresponds to the Siegert relation, $g^{(2)}(\tau)=1+|g^{(1)}(\tau)|^2$, which establishes an equivalence between the (loss of) coherence for the field and the intensity~\cite{Siegert:1943}.

The textbook derivation of the Siegert relation~\cite{Loudon:book} relies on three conditions: a large number of scatterers, the absence of correlations between the emitters (here supported by the negligible interactions), and a zero average electric field $\langle \hat{E}_j\rangle$ for each emitter $j$. The latter condition is provided, for elastic scattering, by thermal motion~\cite{Loudon:book}. For spontaneous emission, the absence of coherence between ground and excited states of each atom $j$ guarantees $\langle \hat E_j^-\rangle=0$ (despite $\langle \hat E_j^- \hat E_j^+\rangle\neq 0$), even in the absence of motion. Yet, even the sum of the two kinds of emission, as encountered in the intermediate $s$ regime, satisfies the Siegert relation, see~\cite{SM}. The reason is that elastically scattered and spontaneously emitted light are uncorrelated fields~\cite{SM}, with independent mechanisms to provide the zero average of the field: Thermal motion relies on the external degrees of freedom of the emitters, whereas spontaneous emission stems from zero-point fluctuations.

{\em Conclusions.---}We have investigated the transition from thermal to spontaneous emission for loss of light coherence in a macroscopic cold atomic cloud using intensity-intensity correlation measurements. This transition is monitored by tuning the pump drive, which controls the ratio between elastic scattering (subjected to thermal broadening) and spontaneous emission. Field-field measurements revealed that no extra dephasing mechanism is present in the electric field, as compared to the intensity.
These results highlight the potential of cold atomic samples to explore the frontier between statistical physics and quantum effects in large systems.

The transition occurs in absence of interactions between the emitters. An open question is how collective effects, arising for example from dipole-dipole interactions (super- and subradiance) may leave a mark on light coherence. The original configuration envisioned for superradiance was the decay cascade from a fully-excited state to the ground state of a many-atom cloud~\cite{Dicke1954}, which is intrinsically an out-of-equilibrium dynamics. Steady-state signatures of collective effects for atoms in free space remain to be explored, which could lead to new phase transitions~\cite{Baumann2010,Bastidas2012,Ferioli2022}. In this context, it is worth mentioning a recent report of steady-state superradiance in intensity correlations measurements in a four-wave mixing experiment~\cite{Araujo2022}.

\begin{acknowledgments}
The authors acknowledge funding from the French National Research Agency (projects
PACE-IN ANR19-QUAN-003-01 and QuaCor ANR19-CE47-0014-01). M.\,A.\,F.\,B., A.\,C. and R.\,B. benefited from Grants from S\~ao Paulo Research Foundation (FAPESP, Grants Nos. 2021/02673-9, 2017/09390-7, 2018/15554-5, 2019/13143-0) and from the National Council for Scientific and Technological Development (CNPq, Grant Nos.\,409946/2018-4 and 313886/2020-2). R.\,K., R.\,B. and M.\,H. received support from the project STIC-AmSud (Ph879-17/CAPES 88887.521971/2020-00).  We also acknowledge the financial support of the Doeblin Federation and R.K. received support from the European project ANDLICA, ERC
Advanced Grant Agreement No. 832219.
\end{acknowledgments}

\bibliography{Biblio}
\pagebreak

\clearpage

\onecolumngrid
\begin{center}
\textbf{\large Supplemental material: \papertitle}\\[.2cm]
\vskip0.5\baselineskip{Pierre Lass\`egues,$^{1}$ Mateus Antônio Fernandes Biscassi,$^{1,2}$ Martial Morisse,$^{1}$ Andr\'e Cidrim,$^{2,3}$ Pablo Gabriel Santos Dias,$^{2}$ Hodei Eneriz,$^{1}$ Raul Celistrino Teixeira,$^{2}$ Robin Kaiser,$^{1}$ Romain Bachelard,$^{1,2}$ and Mathilde Hugbart$^{1}$}
\vskip0.5\baselineskip{{\em$^{1}$Departamento de Física, Universidade Federal de São Carlos,\\ Rodovia Washington Luís, km 235 - SP-310, 13565-905 São Carlos, SP, Brazil}\\
{\em $^{2}$Universit\'e C\^ote d'Azur, CNRS, Institut de Physique de Nice, 06560 Valbonne, France}\\
{\em $^{3}$Department of Physics, Stockholm University, AlbaNova University Center 10691 Stockholm, Sweden}}
\end{center}

\twocolumngrid

\setcounter{equation}{0}
\setcounter{figure}{0}
\setcounter{table}{0}
\setcounter{page}{1}
\makeatletter
\renewcommand{\theequation}{S\arabic{equation}}
\renewcommand{\thefigure}{S\arabic{figure}}
\renewcommand{\bibnumfmt}[1]{[S#1]}
\renewcommand{\citenumfont}[1]{S#1}

\section{Experimental setup}

We here detail the setup presented in Fig.~\ref{fig:setup}(a) of the main text, see Refs.~\cite{Ortiz_2019,Ferreira_2020} for further details. The scattering medium is produced by loading a magneto-optical trap from a vapor of $N\approx 10^8$ $^{85}$Rb atoms, with a low atomic density $\rho\approx 0.005/\lambda^3$ ($\lambda=2\pi/k$ the optical wavelength). After a $2$~ms time of flight, the cloud is illuminated by a flattop intensity laser beam with a frequency $\omega_L$ locked on the $\lvert 3\rangle \rightarrow \lvert 4'\rangle$ hyperfine transition of the D2 line. The beam diameter at the atoms position is $14.7$~mm, which is much larger than the cloud radius ($\sim 0.4$ and $0.8$~mm in the two transverse directions). Hence, the intensity incident on the atoms is uniform (within 10$\,\%$), with Rabi frequency $\Omega$. We use $\lambda/2$ and $\lambda/4$ plates to obtain a circularly polarized light, and the intensity is changed to tune the saturation parameter $s=2\Omega^2/\Gamma^2$ between 0.004 and 60. To maintain similar heating effects over the different regimes, the duration of the laser pulse, always at resonance, is adjusted to get a constant number of photons scattered per atom of $\sim 400$.

The scattered light is collected at $\theta=90^\circ$ from the probe beam axis, using a polarization-maintaining (PM) single-mode fiber. The polarization is selected before the fiber with a $\lambda/2$ plate and a polarization beam splitter (PBS) to maximize the amount of collected photons as well as to adjust the incident polarization along the PM fiber axis. This PM fiber is then connected to a fibered beam splitter (FBS) whose outputs illuminate two single photon counter detectors (avalanche photodiodes APDs) connected to a time-to-digital converter (TDC). The latter device allows to time-tag the arrival of each photon. The second input of the fibered beamsplitter is used to add a local oscillator (LO) derived from the laser which delivers the probe beam. The LO is frequency-shifted by $\omega_\mathrm{BN}=220$~MHz with an acousto-optical modulator (AOM), and its polarization is adjusted before the entrance of the fiber to correspond to the PM fiber axis.

\section{Second-order correlation function for non-interacting atoms}

We derive here the analytical expression for the second-order correlation function at zero delay $g^{(2)}(0)$ for an ensemble of $N$ two-level non-interacting atoms. The detected electric field is assumed to be measured in the far-field and along a direction $\hat n$, so it reads:
\begin{equation}
    \hat E^{+} = E_0 \sum_{a=1}^{N} e^{-ik\hat n\cdot \mathbf{r}_a} \hat\sigma^{-}_a,
\end{equation}
with $\mathbf{r}_a$ the position vector of atom $a$, $\hat\sigma^{\mp}_a$ the two-level lowering/raising spin operation, and $E_0$ a normalization prefactor. Without loss of generality, we hereafter set $E_0=1$, resulting in a normalized electric field intensity which peaks at unity for a single atom. We also assume that the atomic cloud is dilute, thus interaction between the atoms can be disregarded and the steady state of the system is separable. We can then write the state of the system as a direct product as follows
\begin{equation}\label{Eq:SeparableState}
    \hat\rho=\bigotimes_{a=1}^{N}\hat\rho_a,
\end{equation}
where $\hat\rho_a$ is the single-particle density matrix.

\subsection{Scattered field intensity}
Let us first calculate the intensity of the field scattered by the atomic ensemble, which reads
\begin{equation}\label{Eq:IntensityDefition}
\begin{aligned}
I&=\langle \hat E^{-} \hat E^+\rangle=\sum_{ab}\Tr{e^{ik\hat n\cdot \mathbf{r}_a}\hat\sigma_a^{+}e^{-ik\hat n\cdot \mathbf{r}_b}\hat\sigma^{-}_b\hat\rho}\\
&=\sum_{a}\Tr{\hat\sigma_a^{+}\hat\sigma^{-}_a\hat\rho}+\sideset{}{'}\sum_{ab}\Tr{e^{ik\hat n\cdot \mathbf{r}_a}\hat\sigma_a^{+}e^{-ik\hat n\cdot \mathbf{r}_b}\hat\sigma^{-}_b\hat\rho},
\end{aligned}
\end{equation}
where we have introduced the notation $\sideset{}{'}\sum\limits_{a,b\ldots n}\equiv \sideset{}{}\sum\limits_a\sideset{}{}\sum\limits_{b\neq a}\quad\ldots\sideset{}{}\sum\limits_{n\neq a,b\ldots n-1}$.

Using now the separability of the atomic state as in Eq.~\eqref{Eq:SeparableState} and conveniently introducing the excited population 
\begin{equation}\label{Eq:PopulationDefinition}
    n_a \equiv \Tr{\hat\sigma_a^{+}\hat\sigma^{-}_a\hat\rho_a}
\end{equation}
and the coherence
\begin{equation}\label{Eq:CoherenceDefinition}
    \beta_a \equiv e^{-ik\hat n\cdot \mathbf{r}_a}\Tr{\hat\sigma^{-}_a\hat\rho_a}\implies \beta_a^{*} \equiv e^{ik\hat n\cdot \mathbf{r}_a}\Tr{\hat\sigma^{+}_a\hat\rho_a},
\end{equation}
we can rewrite Eq.~\eqref{Eq:IntensityDefition} as
\begin{equation}
\begin{aligned}
I&=\sum_{a}n_a+\sideset{}{'}\sum_{ab}\beta_a^*\beta_b,
\\ &=\sum_{a}n_a+\Big|\sum_{a}\beta_a\Big|^2-\sum_{a}|\beta_a|^2.
\end{aligned}
\end{equation}

\subsection{Unormalized second-order correlation function}

Similarly, following the separability of the atomic state, the second-order correlation reads
\begin{equation}\label{Eq:SecondOrderDefinition}
\begin{aligned}
G^{(2)}&(0)=\langle \hat E^{-} \hat E^{-}\hat E^+ \hat E^+\rangle\\
&= 2\sideset{}{'}\sum_{ab}\Tr{\hat\sigma_a^{+}\hat\sigma^{-}_a\hat\rho_a}\Tr{\hat\sigma_b^{+}\hat\sigma^{-}_b\hat\rho_b}\\
&+4\sideset{}{'}\sum_{abc}\Tr{\hat\sigma_a^{+}\hat\sigma^{-}_a\hat\rho_a}\Tr{e^{ik\hat n\cdot \mathbf{r}_b}\hat\sigma^{+}_b\hat\rho_b}\\
&\quad\quad\quad\,\,\,\,\Tr{e^{-ik\hat n\cdot \mathbf{r}_c}\hat\sigma^{-}_c\hat\rho_c}\\
&+\sideset{}{'}\sum_{abcd}\Tr{e^{ik\hat n\cdot \mathbf{r}_a}\hat\sigma_a^{+}\hat\rho_a}\Tr{e^{ik\hat n\cdot \mathbf{r}_b}\hat\sigma^{+}_b\hat\rho_b}\\
&\quad\quad\quad\,\Tr{e^{-ik\hat n\cdot \mathbf{r}_c}\hat\sigma_c^{-}\hat\rho_c}\Tr{e^{-ik\hat n\cdot \mathbf{r}_d}\hat\sigma^{-}_d\hat\rho_d}.
\end{aligned}
\end{equation}
Using the definitions in Eqs.~\eqref{Eq:PopulationDefinition} and \eqref{Eq:CoherenceDefinition}, we are left with
\begin{equation}
G^{(2)}(0)=2\sideset{}{'}\sum_{ab}n_an_b+4\sideset{}{'}\sum_{abc}n_a\beta_b^*\beta_c+\sideset{}{'}\sum_{abcd}\beta_a^{*}\beta_b^{*}\beta_c\beta_d.
\end{equation}
Reorganizing the expression using sums without index exclusion, one can expand the expression above as
\begin{equation}\label{Eq:G2Separable}
\begin{aligned}
G^{(2)}(0)=&2\Big(\sum_a n_a\Big)^2-2\sum_a n_a^2\\ 
+&4\Big(\sum_a n_a\Big)\Big(\Big|\sum_b\beta_b\Big|^2-\sum_b|\beta_b|^2\Big)\\
-&8\mathrm{Re}\Big\{\Big(\sum_a n_a\beta^*_a\Big)\Big(\sum_b \beta_b\Big)\Big\}+8\sum_a n_a |\beta_a|^2\\
+&\Big|\sum_a \beta_a\Big|^4-6\sum_a |\beta_a|^4 - 4\Big|\sum_a \beta_a\Big|^2\Big(\sum_b |\beta_b|^2\Big)\\
+&8\mathrm{Re}\Big\{\Big(\sum_a \beta_a\Big)\Big(\sum_b |\beta_b|^2\beta_b^*\Big)\Big\}+2\Big(\sum_a |\beta_a|^2\Big)^2\\
-&2\mathrm{Re}\Big\{\Big(\sum_a \beta_a\Big)^2\Big(\sum_b (\beta^*_b)^2\Big)\Big\}+\Big|\sum_a \beta^2_a\Big|^2.
\end{aligned}
\end{equation}

\subsection{Separable steady-state as a function of the saturation parameter}

Considering a laser with wave-vector $\mathbf{k}_L$ driving a two-level atom on its resonance at a Rabi frequency $\Omega$, the single-atom density matrix in the steady state is given by
\begin{equation}
\begin{aligned}
\hat\rho_a=&\rho^{(a)}_{ee}|e\rangle\langle e|+\rho^{(a)}_{eg}|e\rangle\langle g|\\ +&\rho^{(a)}_{ge}|g\rangle\langle e|+\rho^{(a)}_{gg}|g\rangle\langle g|,
\end{aligned}
\end{equation}
with 
\begin{equation}
\begin{aligned}
\rho^{(a)}_{ge}&=\Big(\rho^{(a)}_{eg}\Big)^*=-i\frac{e^{-i\mathbf{k}_L.\mathbf{r}_a}}{1+s}\sqrt{\frac{s}{2}},\\ \rho^{(a)}_{ee}&=\frac{s}{2(1+s)},\\
\rho^{(a)}_{gg}&=\frac{2+s}{2(1+s)},
\end{aligned}    
\end{equation}
where $s\equiv 2\Omega^2/(\Gamma^2+4\Delta^2)$ is the saturation parameter, which on resonance ($\Delta=0$) reduces to the ratio $s=2\Omega^2/\Gamma^2=P_\mathrm{SE}/P_\mathrm{EL}$, as discussed in the main text.

Substituting the elements of the single-particle density matrix in the definition of $g^{(2)}(0)=G^{(2)}(0)/I^2$, one is left with 
\begin{widetext}
\begin{equation}
\begin{aligned}
g^{(2)}(0)=&\frac{1}{(N s + |\Phi_{1}|^2)^2}\Bigg(2Ns[2+(N-1)s]+4s(N-2)|\Phi_1|^2+|\Phi_1^2-\Phi_2|^2\Bigg),
\end{aligned}
\end{equation}
\end{widetext}
where we have defined 
\begin{equation}
\begin{aligned}
\Phi_1&=\sum_a e^{i(k\hat n-\mathbf{k}_L)\cdot \mathbf{r}_a},\\
\Phi_2&=\sum_a e^{i2(k\hat n -\mathbf{k}_L)\cdot \mathbf{r}_a}.
\end{aligned}
\end{equation}
For $s\to\infty$, one recovers the formula $g^{(2)}(0)=2(1-1/N)$. In the limit $s\to\infty$, the destructive interference condition $\Phi_1=0$ leads to a $g^{(2)}(0)$ scaling as $(|\Phi_2|/sN)^2$, which diverges for $s\to 0$ at fixed $N$.

\section{Siegert relation for two independent fields}

Here we show in a condensed manner that the sum of two fields, each satisfying the Siegert relation and the associated conditions described in the main text, also satisfies the relation, provided that the fields are uncorrelated. The derivation is provided for two arbitrary electric fields $\hat E^{+}_e$ and $\hat E^{+}_i$ corresponding, for example, to the elastically and inelastically (that is, spontaneously emitted) electric fields of the main text. The total electric field is $\hat{E}^+ = \hat{E}_e^+ + \hat{E}_i^+$, and it presents the following second-order correlation function in the steady state:
\begin{equation}
\begin{aligned}
    G^{(2)}(\tau)= & \langle [\hat{E}^{-}_e(t) + \hat{E}^{-}_i(t)] [\hat{E}^{-}_e(t+\tau) + \hat{E}^{-}_i(t+\tau)]\\
    \times&[\hat{E}^{+}_e(t+\tau) + \hat{E}^{+}_i(t+\tau)] [\hat{E}^{+}_e(t) + \hat{E}^{+}_i(t)]\rangle.
\end{aligned}
\end{equation}
Both elastic and inelastic terms of the electric field have a zero average: $\langle \hat{E}_{e,i}\rangle=0$, so their sum as well: $\langle \hat{E}\rangle=0$. Furthermore, the absence of correlation between the fields makes that the contributions from the elastic and inelastic terms can be factorized in the above expression of $G^{(2)}$, which in turn leads to the cancellation of several terms:
\begin{equation}
\begin{aligned}
    G^{(2)}(\tau)= & \langle \hat{E}^{-}_e(t) \hat{E}^{-}_e(t+\tau)  \hat{E}^{+}_e(t+\tau) \hat{E}^{+}_e(t) \rangle  \\
    +&\langle \hat{E}^{-}_i(t) \hat{E}^{-}_i(t+\tau)  \hat{E}^{+}_i(t+\tau) \hat{E}^{+}_i(t) \rangle\\
    +&2\langle \hat{E}^{-}_e(t) \hat{E}^{+}_e(t)\rangle\langle \hat{E}^{-}_i(t) \hat{E}^{+}_i(t)\rangle\\
    +&2\mathrm{Re}\left[\langle \hat{E}^{-}_e(t+\tau) \hat{E}^{+}_e(t)\rangle\langle \hat{E}^{-}_i(t) \hat{E}^{+}_i(t+\tau) \rangle\right].
\end{aligned}
\end{equation}
Considering now that each scatterer presents the same single-particle correlation functions $G^{s(1)}_{e,i}(\tau)$ and contributes equally to the total field, then the single emitter correlation function of the field writes $G^{s(1)}(\tau)=G^{s(1)}_{e}(\tau)+G^{s(1)}_{i}(\tau)$, where we have used that the fields are uncorrelated and have zero average. The correlation function of the total field is then equal to $\langle \hat{E}^{-}(t)\hat{E}^{-}(t+\tau)\rangle=NG^{s(1)}(\tau)$, with $N$ the number of scatterers. Calling $G^{(2)}_{e,i}(\tau)$ the non-normalized second order correlation function of the total elastic or inelastic terms, respectively, we obtain:
\begin{equation}
\begin{aligned}
    G^{(2)}(\tau) =& G^{(2)}_{e}(\tau)+G^{(2)}_{i}(\tau)+2N^2G^{s(1)}_{e}(0)G^{s(1)}_{i}(0)\\
     +& 2N^2\mathrm{Re}\left[G^{s(1)}_{e}(\tau)G^{s(1)*}_{i}(\tau)\right].\label{eq:G2a}
\end{aligned}
\end{equation}
Assuming that the spectrum emitted by a single atom is symmetric, its Fourier transform $G^{s(1)}_{e,i}(\tau)$ is real. We also assumed that both the elastic and inelastic fields satisfy the Siegert relation,  so \eqref{eq:G2a} can be rewritten as:
\begin{equation}
\begin{aligned}
    G^{(2)}(\tau)=& N^2\left( \left[G^{s(1)}_{e}(0)\right]^2 +\left[G^{s(1)}_{e}(\tau)\right]^2 \right)\\
   +& N^2\left( \left[G^{s(1)}_{i}(0)\right]^2 +\left[G^{s(1)}_{i}(\tau)\right]^2 \right)\\
    +& 2N^2G^{s(1)}_{e}(0)G^{s(1)}_{i}(0)+ 2N^2G^{s(1)}_{e}(\tau)G^{s(1)}_{i}(\tau).
\end{aligned}
\end{equation}
This can be simplified to:
\begin{equation}
\begin{aligned}
    g^{(2)}(\tau)&= \frac{\left[G^{s(1)}_{e}(0)+G^{s(1)}_{i}(0)\right]^2 + \left[G^{s(1)}_{e}(\tau)+G^{s(1)}_{i}(\tau)\right]^2} {\left[G^{s(1)}_{e}(0)+G^{s(1)}_{i}(0)\right]^2} \\ &= 1 + |g^{(1)}(\tau)|^2.
\end{aligned}
\end{equation}
Thus, the sum of two uncorrelated Siegert-satisfying field still satisfies the Siegert relation. For our particular system, it implies that in the intermediate saturation regime, where both elastic and inelastic scattering occur, the Siegert relation is verified for large number of scatterers.

\end{document}